\documentclass[twocolumn,amsmath,amssymb,aps,longbibliography,floatfix]{revtex4-1}

\usepackage{cancel}
\usepackage{enumitem}
\usepackage[utf8x]{inputenc}
\usepackage{graphicx}
\usepackage{dcolumn}
\usepackage{bm}
\usepackage{hyperref}
\usepackage[switch]{lineno}
\usepackage{float}             
\usepackage{subfigure}
\usepackage{tikz}

\begin{document}

\title{Spin-Phonon Relaxation in Molecular Qubits from First Principles Spin Dynamics}

\author{$^{1}$Alessandro Lunghi}
\email{lunghia@tcd.ie}
\author{$^{1}$Stefano Sanvito}

\affiliation{$^{1}$School of Physics, AMBER and CRANN, Trinity College, Dublin 2, Ireland}

\begin{abstract}
{\bf The coupling between electronic spins and lattice vibrations is fundamental for driving relaxation in magnetic materials. 
The debate over the nature of spin-phonon coupling dates back to the 40's, but the role of spin-spin, spin-orbit and 
hyperfine interactions, has never been fully established. Here we present a comprehensive study of the spin dynamics of a 
crystal of Vanadyl-based molecular qubits by means of first-order perturbation theory and first-principles calculations. We 
quantitatively determine the role of the Zeeman, hyperfine and electronic spin dipolar interactions in the direct mechanism 
of spin relaxation. We show that, in a high magnetic field regime, the modulation of the Zeeman Hamiltonian by the intra-molecular 
components of the acoustic phonons dominates the relaxation mechanism. In low fields, hyperfine coupling takes over, with the 
role of spin-spin dipolar interaction remaining the less important for the spin relaxation.}
\end{abstract}

\maketitle

Spin 1/2 systems represent the fundamental prototype of magnetic materials and the understanding of their properties 
is pivotal for the rationalization of any more complex magnetic compound. Their study has deep roots in the early days of 
quantum mechanics and, despite their basic nature, they still represent a very rich quantum playground with non-trivial and elusive 
dynamical properties. The spin life-time in insulating materials is essentially limited by the interaction between spins and lattice 
vibrations, namely the spin-phonon coupling. At finite temperature, spins can absorb/emit one or multiple phonons from/into the 
lattice and relax to an equilibrium state. The detailed understanding of this process at the first-principles level represents a 
long-standing question in physics and chemistry and goes well beyond the fundamental-theory aspect. 

Spin-lattice relaxation is key in several fields. For instance, the efficiency of magnetic-resonance-imaging contrast 
agents~\cite{Caravan2006} and the spin life-time of single molecule magnets~\cite{Sessoli1993} is determined by the 
magnetization decay rate of paramagnetic elements. Turning to the main focus of this work, in both molecular and solid-state 
qubits, spin-lattice relaxation sets the upper limit for the coherence time~\cite{Balasubramanian2009,Zadrozny2015,Bader2016,Whiteley2019} 
and the engineering of this interaction is a primary challenge in the spin-based quantum computing field. The synthetic 
versatility of molecular compounds offers an intriguing route to the tailoring of spin-phonon coupling but needs to be supported 
by a rational understanding of the physical principles governing spin relaxation.

The debate over the role of phonons in the relaxation of electronic spins can be traced back to the 40's, 
when it was discussed for the first time in the context of transition-metal chemistry. Relaxation pathways 
occurring through the modulation of spin dipolar interactions or the modulation of the $d$-electrons' crystal field 
were firstly pointed out by Waller~\cite{waller1936waller} and Van Vleck~\cite{Vleck1940,Orbach1961}, respectively.
More recently, the role of hyperfine interaction has also been discussed~\cite{Gomez-Coca2014}.
Van Vleck's mechanism remains to date the most commonly accepted explanation for the microscopic origin of 
spin-lattice relaxation of electronic spins. Several questions, however, remain unanswered. In particular, early 
models, being phenomenological, could not entirely address the nature of the coupling between 
phonons and spins and simply ascribe acoustic vibrations as responsible for this interaction.
Molecular compounds are inherently complex and a rational description of spin relaxation in terms of molecular motions is 
still to be developed.

First-principles calculations represent the perfect ground to provide unbiased answers to such fundamental 
questions. The possibility to predict the spin-relaxation times in magnetic materials without 
introducing any phenomenological parameter is also of fundamental importance from a materials-design 
perspective. Quantum chemistry and related disciplines have been already proved to be invaluable tools for 
providing insights into the physics of new chemical systems~\cite{Neese2019} and they have been used as 
a screening method to predict new materials with tailored properties~\cite{Curtarolo2013}. These computational 
strategies represent a rich opportunity for the field of quantum informatics. The development of a first-principle 
framework able to predict the spin-relaxation time is the first step in that direction. In this work we offer a theoretical 
and computational explanation of the nature of the atomistic processes that lead to the relaxation of a molecular 
electronic $S=1/2$ system. 

Our approach builds on previous contributions from ourselves~\cite{Lunghi2017,Lunghi2017a} and 
others~\cite{Escalera-Moreno2017,Goodwin2017,Astner2018,Moseley2018} 
and significantly extends the state of the art in the field. For the first time we include in the formulation the contribution
of phonons from the entire Brillouin zone and the hyperfine and dipolar spin-spin interactions, both at a static and 
dynamical level. The description of all the fundamental spin-relaxation channels occurring at the first-order of 
perturbation theory is therefore here complete. Our method is purely first-principle and provides a 
rationalization of spin-lattice relaxation without any previous knowledge of the system's properties other than 
its crystal structure. The $S=1/2$ system investigated is the crystal of the Vanadyl complex VO(acac)$_{2}$, 
being acac=acetylacetonate~\cite{Tesi2016}. Vanadyl-based molecular qubits represent archetypal systems for 
room-temperature quantum-computing applications~\cite{Atzori2016}.

We demonstrate that Van Vleck's mechanism is the dominant first-order relaxation channel for $S=1/2$ molecular spins 
in a high magnetic field, while hyperfine contributions become dominant in a low-field regime. Most importantly, our calculations 
show that acoustic phonons are not rigid molecular translations, shedding light on the origin of spin-phonon coupling in 
solid-state spin 1/2 compounds, and thus solving a eighty-year-old controversy.

\section{Results}

\subsection{First-Principles Spin Dynamics}

The quantum dynamics of even a single spin is, in principle, entangled with the dynamics of all the other spin and 
lattice degrees of freedom that it is interacting with. More explicitly, it is driven by the total Hamiltonian 
$H=H_\mathrm{s}+H_\mathrm{ph}+H_\mathrm{sph}$, where the three terms are, respectively, the spins and phonons 
Hamiltonian, and the spin-phonon coupling. It is convenient to think at the problem as composed of two parts: the 
simulation of the sole spin degrees of freedom as an isolated system and the interaction of this ensemble with a thermal 
bath, namely with the phonons. 

The spin Hamiltonian of Eq.~(\ref{SH}) describes the fundamental interactions taking place within an ensemble of $N_\mathrm{s}$ 
spins,
\begin{equation}
\hat{H}_\mathrm{s}=\sum_{i}^{N_\mathrm{s}}\beta_{i}\vec{\mathbf{B}}\cdot\mathbf{g}(i)\cdot\vec{\mathbf{S}}(i)+\frac{1}{2}\sum_{ij}^{N_{s}}\vec{\mathbf{S}}(i)\cdot\mathbf{D}(ij)\cdot\vec{\mathbf{S}}(j)\:,
\label{SH}
\end{equation}
where the $i$-th spin, $\mathbf{S}(i)$, interacts with an external magnetic field $\vec{\mathbf{B}}$ through the 
gyromagnetic tensor, $\beta\mathbf{g}(i)$. Eq.~(\ref{SH}) can account for both electronic and nuclear spins on the 
same footing. Thus the spin-spin interaction tensor, $\mathbf{D}$, may coincide with the point-dipole interaction, 
$\mathbf{D}^\mathrm{dip}$, or the hyperfine tensor, $\mathbf{A}$, depending on whether the interaction is among 
electronic and/or nuclear spins. The state of the spin system can be described in term of the spin density matrix, 
$\hat{\rho}$, whose dynamics is regulated by the Liouville equation,
\begin{equation}
\frac{d\hat{\rho}}{dt}=-\frac{i}{\hbar}[\hat{H}_\mathrm{s},\hat{\rho}]\:.
\label{UN}
\end{equation} 

Once the eigenstates and eigenvalues, $\{E_{a}\}$, of the spin Hamiltonian are known, it is possible to 
integrate exactly Eq.~(\ref{UN}) and obtain an expression for the time evolution of $\rho^\mathrm{I}_{ab}$ 
in the so-called configuration interaction,
\begin{equation}
\rho^\mathrm{I}_{ab}(t)=e^{-i\omega_{ab}(t-t_{0})}\rho_{ab}(t_{0})\:,
\label{UN1}
\end{equation}
where $\omega_{ab}=(E_{a}-E_{b})/\hbar$. Despite its simple form, Eq.~(\ref{UN1}) hides a high level of 
complexity, which originates from the high dimension of the Hilbert space it acts on. Several numerical 
schemes exist to reduce the computational costs associated to this problem~\cite{Daley2004,Kuprov2007} 
and these will be the subject of future investigations. Here we have decided to retain the full Hilbert space 
description and restrict the study to a small number of interacting spins. 

The basic theory for phonon-driven spin relaxation has been derived before~\cite{Lunghi2017,Lunghi2017a}. 
Here we extend it by including acoustic phonons and the phonon reciprocal space dispersion. For a periodic 
crystal, defined by a set of reciprocal lattice vectors, $\mathbf{q}$, and by $N$ atoms in the unit-cell, the 
lattice's dynamics can be described in terms of periodic displacement waves (phonons), $Q_{\alpha\mathbf{q}}$, 
with frequency $\omega_{\alpha\mathbf{q}}$ and obeying to an harmonic Hamiltonian
\begin{equation}
\hat{H}_\mathrm{ph}=\sum_{\alpha\mathbf{q}}\hbar\omega_{\alpha\mathbf{q}}(\hat{n}_{\alpha\mathbf{q}}+1)\:.
\label{HPH}
\end{equation}
Here $\hat{n}_{\alpha\mathbf{q}}$ is the phonon's number operator. The spin-phonon coupling Hamiltonian, 
responsible for the energy exchange between the spins and the lattice, writes
\begin{equation}   
H_\mathrm{sph}=\sum_{\alpha\mathbf{q}}\Big(\frac{\partial H_{s}}{\partial Q_{\alpha\mathbf{q}}}\Big)Q_{\alpha\mathbf{q}}\:.
\label{SPH}
\end{equation}
Usually, phonons dynamics develops at a time scale much shorter than that of the spin relaxation. Therefore, if no 
phonon bottleneck is at play, the Born-Markov approximation will be valid and the phonons dynamics can be 
considered to be always at the thermal equilibrium. Under these circumstances the full set of Redfield equations 
can be used to study the spin dynamics under the effect of a phonon bath~\cite{Petruccione2002}. In this framework, 
the spin density matrix $\rho_{ab}^\mathrm{I}(t)$ evolves according to the equation,
\begin{equation}
\frac{d\rho^\mathrm{I}_{ab}}{dt}=\sum_{cd}\sum_{\alpha\mathbf{q}}R^{\alpha\mathbf{q}}_{ab,cd}\rho^\mathrm{I}_{cd}(t)\:.
\label{RED}
\end{equation}
The transition rates between the elements of the density matrix are of the form  
\begin{equation}
R^{\alpha\mathbf{q}}_{ab,cd}\propto\frac{\pi}{2\hbar^2}V^{\alpha\mathbf{q}}_{ac}V^{\alpha\mathbf{q}}_{db}
\Big(G(\omega_{db},\omega_{\alpha\mathbf{q}})+G(\omega_{cd},\omega_{\alpha\mathbf{q}})\Big)\:,
\label{RED0}
\end{equation}
where $V_{ab}^{\alpha\mathbf{q}}=\langle a|\frac{\partial H_{s}}{\partial Q_{\alpha\mathbf{q}}}|b\rangle$ is the matrix 
element of the spin-phonon Hamiltonian in the spin Hamiltonian eigenfunctions basis and 
$G(\omega_{ij},\omega_{\alpha\mathbf{q}})$ is the single phonon correlation function at finite temperature. 
For harmonic lattices, $G(\omega_{ij},\omega_{\alpha\mathbf{q}})$ is defined as
\begin{equation}
G(\omega_{ij},\omega_{\alpha\mathbf{q}})= \bar{n}_{\alpha\mathbf{q}}\delta(\omega_{\alpha\mathbf{q}}-\omega_{ij})+(\bar{n}_{\alpha\mathbf{q}}+1)\delta(\omega_{\alpha\mathbf{q}}+\omega_{ij})\:,
\label{delta}
\end{equation}
where $\bar{n}_{\alpha\mathbf{q}}$ is the Bose-Einstein population at a temperature $T$. Once Eq.~(\ref{RED}) is 
solved, the magnetization dynamics for the $i$-th spin can the computed by the canonical expression 
$\vec{\mathbf{M}}(i)=\mathrm{Tr}\{\hat{\rho}(t)\vec{\mathbf{S}}(i)\}$.

\subsection{Spin Phonon Coupling in Molecular Qubits}

\begin{figure*}
\includegraphics[scale=1]{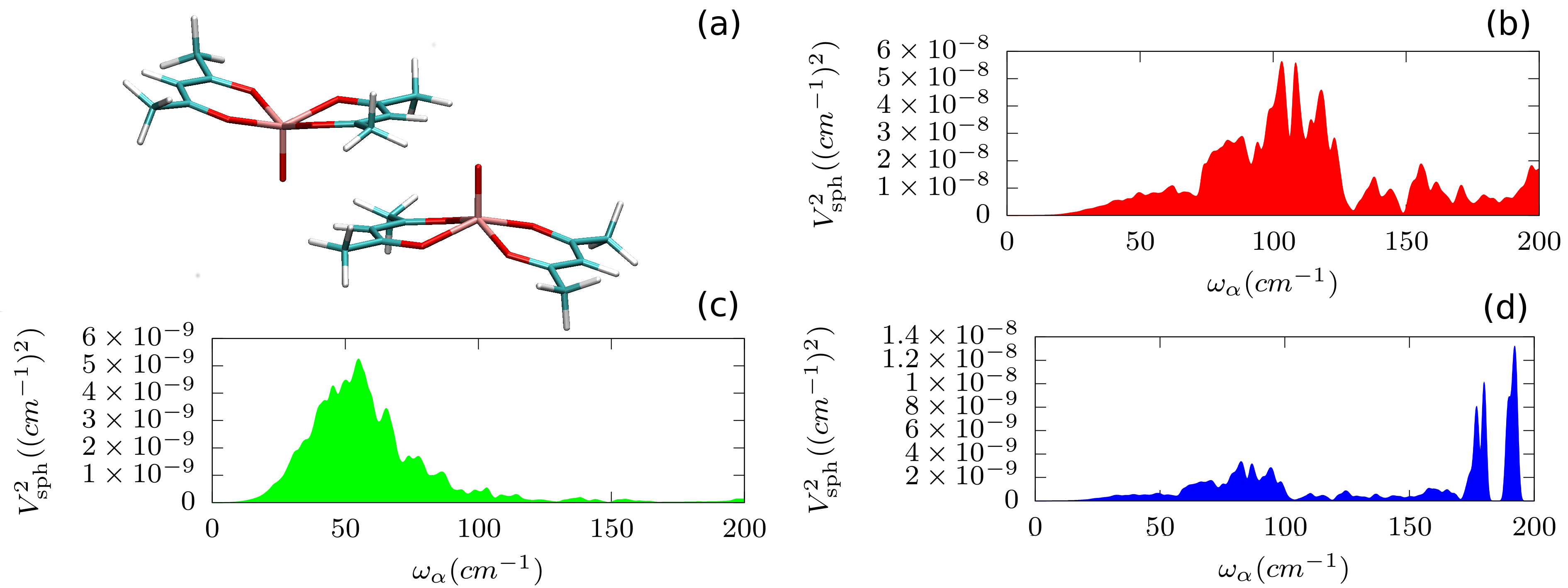}
\caption{\textbf{VO(acac)$_{2}$ structure and spin phonon coupling distributions.} Panel (a) shows the 
geometrical structure of the two VO(acac)$_{2}$ molecular units inside the crystal's unit-cell. Vanadium 
atoms are represented in pink, oxygen in red, carbon in green and hydrogen in white. Panel (b) shows the 
spin-phonon coupling distribution relative to the Zeeman energy as function of the phonons' frequency. 
Panel (c) shows the spin-phonon coupling distribution relative to the dipolar spin-spin energy as function 
of the phonons' frequency. Panel (d) shows the spin-phonon coupling distribution relative to the hyperfine 
energy as function of the phonons' frequency.}
\label{sph_dist}
\end{figure*} 

In order to make the physics of Eq.~(\ref{RED0}) more transparent, a study of the spin-phonon 
coupling terms, $V_{ab}$, for the molecular qubit VO(acac)$_{2}$ is hereafter provided. This V$^{4+}$ complex 
crystallises with a triclinic primitive cell containing two inversion-symmetry-related molecular units~\cite{Tesi2016}, 
as reported in Fig.~\ref{sph_dist}. Each molecule bears a single electronic $S=1/2$ spin in addition to the $I=7/2$ 
nuclear spin of $^{51}$V. Let us consider one electronic spin $\vec{\mathbf{S}}_{i}$ interacting with the rest of the 
crystal electronic spins $\vec{\mathbf{S}}_{j}$ and the 
local vanadium nuclear spin $\vec{\mathbf{I}}_{i}$. Given the definition in Eq.~(\ref{SPH}), the spin-phonon 
coupling Hamiltonian will contain three distinguished contributions: a first intra-molecular spin-phonon coupling 
due to the modulation of the Land\'e tensor, $\mathbf{g}$; a second intra-molecular coupling coming from the 
modulation of the hyperfine interaction, $\mathbf{A}$; and an inter-electronic spins interaction originating from 
the modulation of the dipolar terms, $\mathbf{D}^\mathrm{dip}$, namely
\begin{equation}
\begin{aligned}
\frac{\partial H_\mathrm{s}(i)}{\partial Q_{\alpha\mathbf{q}}}=&\beta_{i}\vec{\mathbf{B}}\cdot\frac{\partial\mathbf{g}(i)}{\partial Q_{\alpha\mathbf{q}}}\cdot\vec{\mathbf{S}}(i)
+\vec{\mathbf{S}}(i)\cdot\frac{\partial\mathbf{A}(ii)}{\partial Q_{\alpha\mathbf{q}}}\cdot\vec{\mathbf{I}}(i)+\\
+&\sum_{j}^{N_{s}}\vec{\mathbf{S}}(i)\cdot\frac{\partial\mathbf{D}^\mathrm{dip}(ij)}{\partial Q_{\alpha\mathbf{q}}}\cdot\vec{\mathbf{S}}(j)\:.
\label{spc}
\end{aligned}
\end{equation}

A 3$\times$3$\times$3 super-cell containing 1620 atoms is optimised at the density functional theory (DFT) level 
and it is used to compute the lattice vibrational properties. The molecular optimised structure has been further employed 
for the calculation of all the spin-phonon coupling coefficients appearing in Eq.~(\ref{spc}). The details 
about the protocol used for the phonons and spin-phonon calculations can be found in the Method Section. All 
these interactions break the single-spin time-reversal symmetry and are potentially active in intra-Kramer-doublet 
spin relaxation, but the physics beyond the three processes is quite different. In order to make these differences 
more evident we have calculated the spin-phonon coupling squared norms $V_\mathrm{sph}^{2}$ (defined in the 
Methods section) associated to each phonon modes and their corresponding distributions. The results, reported in 
panels (b) through (d) of Fig.~\ref{sph_dist}, show striking differences both in qualitative and quantitative terms. 

The anisotropy of the spin Hamiltonian is the fingerprint of the dependence of spin degrees of freedom on atomic 
positions and the same interactions contributing to magnetic anisotropy are also contributing to the 
derivatives of Eq. \ref{spc}. The tensors $\mathbf{g}$ and $\mathbf{A}$ have anisotropic components coming from 
spin-orbit coupling and other interactions localized on the Vanadium centre. Such short-ranged interactions are only 
influenced by localized intra-molecular vibrations and local rotations affecting the first coordination sphere. In this case, 
high-energy phonons are still operative. Conversely, dipolar interactions act between different molecules and are 
both non-local and long-wavelength in nature. These are expected to be prominently modulated by molecular 
translations. Their interaction thus vanishes at high frequency.

\subsection{VO(acac)$_{2}$ Crystal Spin Dynamics}

The spin-phonon coupling coefficients discussed in the previous section have been used together with 
Eq.~(\ref{RED}) to simulate the spin dynamics of three different systems: one single electronic spin, one 
electronic spin interacting with the V nuclear spin and two interacting electronic spins. Tests including more 
that two coupled electronic spins are reported in the supplementary information (SI) and show a very small 
dependence of the relaxation time, $\tau$, on spins farther than the first-neighbour ones. In all simulations, 
spins are interacting with all the phonons of the periodic crystal, calculated by integrating the Brillouin zone 
with homogeneous grids up to 64$\times$64$\times$64 $k$-points. 

\begin{figure}[h!]
\includegraphics[scale=1]{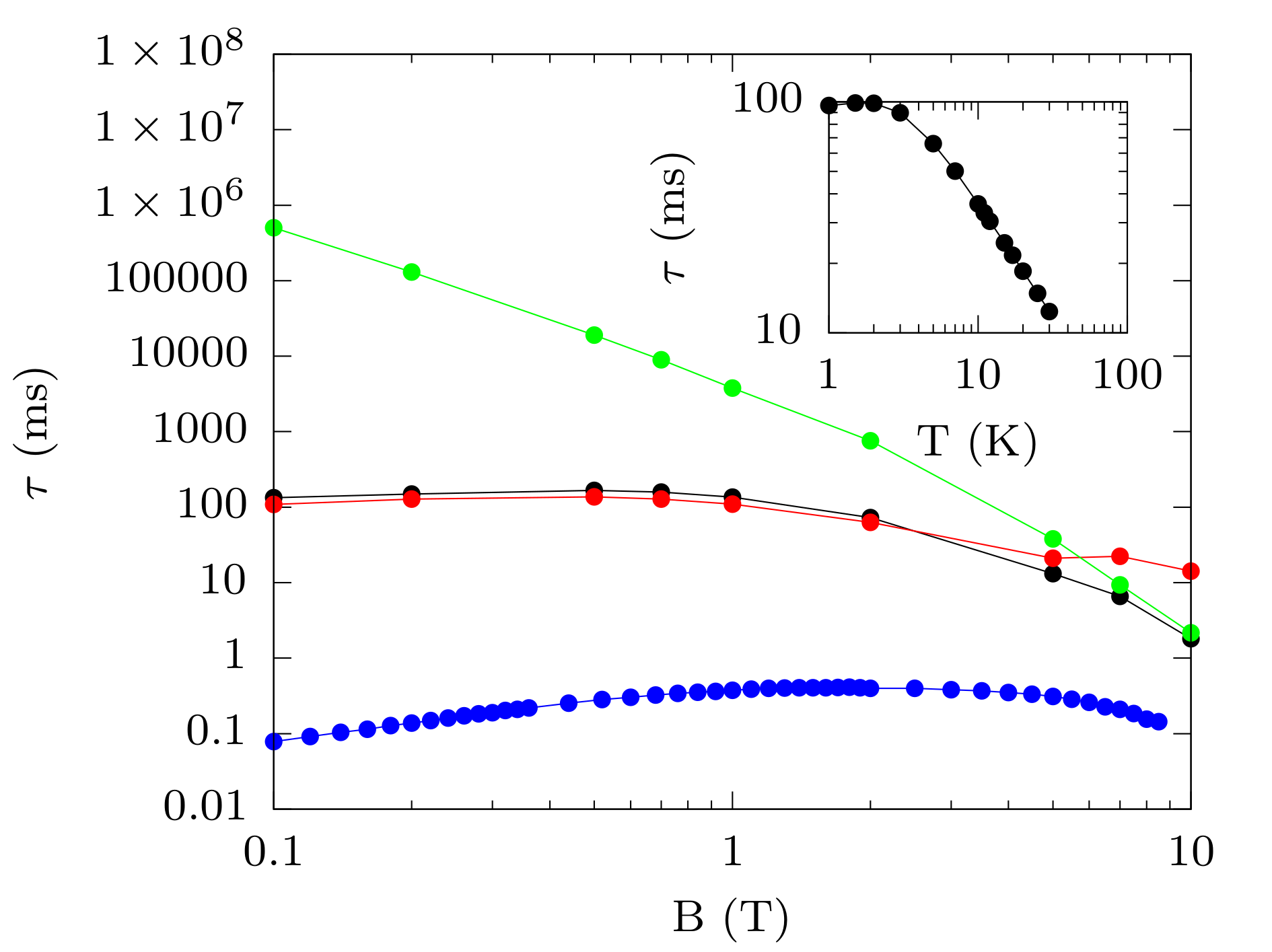}
\caption{\textbf{Spin Relaxation time as function of the B field and the temperature for one electronic spin 
coupled to one nuclear spin.} The relaxation time, $\tau$, in ms as a function of the external field in Tesla is 
reported for the simulations of one electronic spin coupled to one nuclear spin and relaxing due to the 
phonons modulation of the Zeeman and hyperfine energies (black line and dots). The contribution coming 
from the sole hyperfine energy modulation is reported with a red line and dots, while the sole Zeeman contribution 
is reported by the green line and dots. The experimental relaxation time as extracted from AC magnetometry~\cite{Tesi2016}
is also reported (blue dots and line). The inset describes the simulated temperature dependence 
of spin relaxation at 5 Tesla.}
\label{tauB}
\end{figure}

Fig.~\ref{tauB} shows the spin relaxation time as function of the external magnetic field $\vec{\mathbf{B}}$ 
at 20~K for a single electronic spin coupled to the V nuclear spin. The combined effect of the Zeeman and 
hyperfine modulation, described by the black line and dots, shows a field dependence approaching $|\vec{\mathbf{B}}|^4$ 
in the high field limit ($B > 5$~T), where $\tau$ starts to converge to the experimental value~\cite{Tesi2016}. The 
green curve in Fig.~\ref{tauB}, relative to the single electronic spin contribution, shows that in this regime the 
leading mechanism is the modulation of the Zeeman energy. In the low-field regime the relaxation time as a function
of field approaches a plateau where it is about three orders of magnitude longer than the experimental value. 
The red curve in Fig.~\ref{tauB} shows that the rate-determining mechanism in this regime is the modulation 
of the hyperfine Hamiltonian. 

Numerical tests concerning the effects of the numerical noise in the calculation of the frequencies and the spin-phonon 
coupling parameters are reported in the SI and prove the robustness of our results. However, it is important to remark 
that the energies at play in the low-field regime ($B < 1$~T) are extremely small, if compared with the phonons frequencies 
and the simulations. Thus, a more sophisticated Brillouin-zone integration scheme might be needed to obtain a more robust 
estimate of the relaxation times in this field regime. Nonetheless, the calculated relaxation time for low fields is orders of 
magnitude longer than the experimental one, suggesting the presence of higher-order relaxation mechanisms at play in this 
regime.

The inset of Fig.~\ref{tauB} displays the temperature dependence of the spin relaxation time in a field of 5~T. 
In contrast to the experimental results, which show $\tau\propto T^{-n}$ (with $n>2$)~\cite{Tesi2016}, in
this field range we simulate a T$^{-1}$ behaviour. Our result is in agreement with what expected from a first-order approximation to spin-phonon coupling and a harmonic lattice dynamics. Deviations from the T$^{-1}$ power law can be considered as fingerprints of higher-order processes taking place.
Interestingly, our simulations also show a residual $T$-independent process in the $T\rightarrow 0$ limit. In this regime, 
due to the absence of populated phonons states, the only possible relaxation pathway is provided by the $T$-independent 
spontaneous phonons emission from a spin excited state. This phenomenon has been recently observed in N-V 
centres~\cite{Astner2018}.

\begin{figure}[h!]
\includegraphics[scale=1]{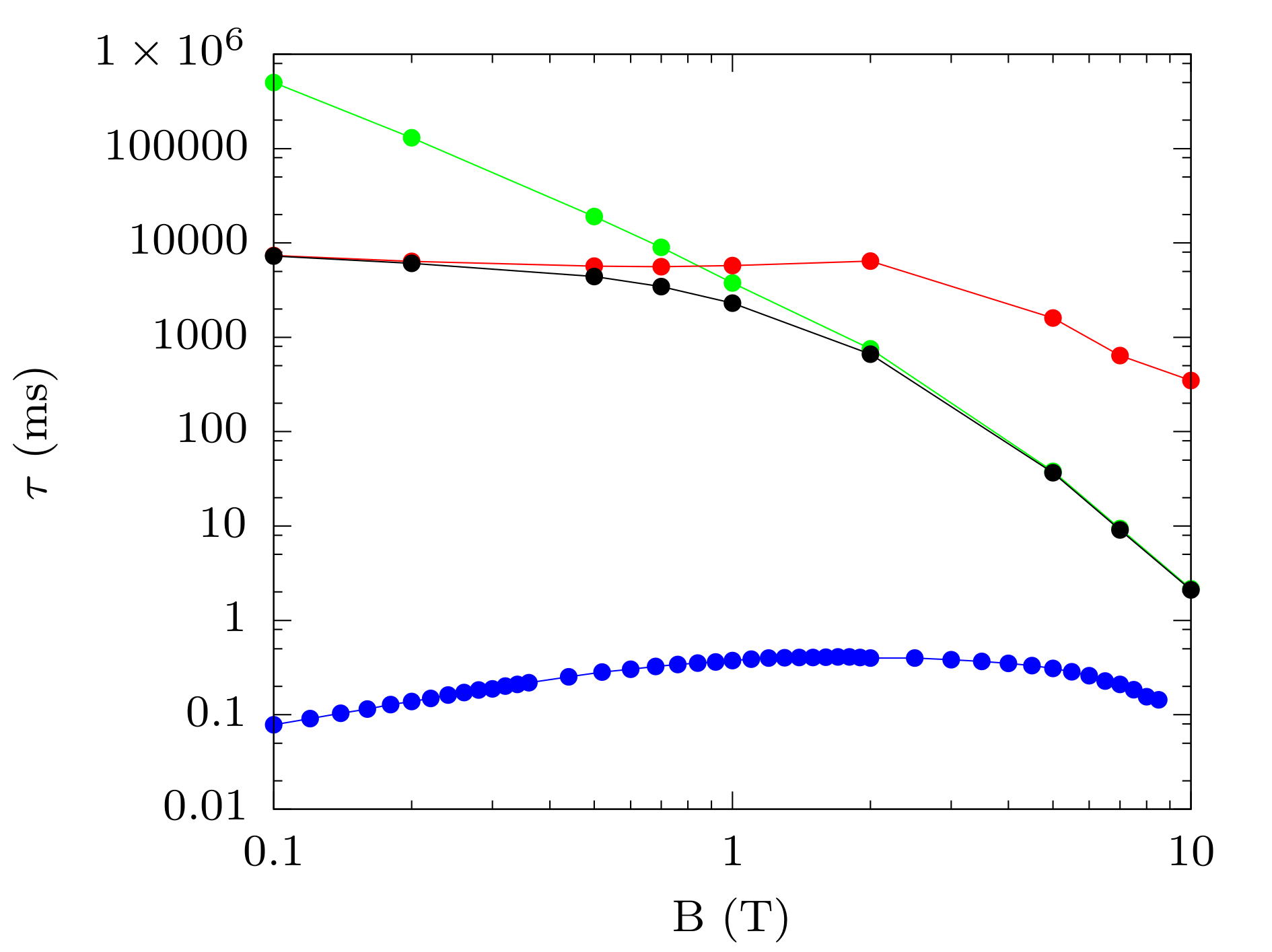}
\caption{\textbf{Spin relaxation time as function of external field for two coupled electronic spins.}
The relaxation time, $\tau$, in ms as a function of the external field in Tesla is reported for the simulations of 
two electronic spins relaxing due to the phonons modulation of the Zeeman and dipolar energies (black line 
and dots). Green line and dots represent the relaxation time of two isolated spins, where only the Zeeman energy 
is modulated by phonons. The contribution coming from the sole dipolar energy modulation is reported with red line 
and dots. The experimental relaxation time as extracted from AC magnetometry~\cite{Tesi2016}
is also reported (blue dots and line).}
\label{tauB2}
\end{figure} 

Fig.~\ref{tauB2} presents the spin-relaxation time as function of the external magnetic field at 20~K for two 
coupled electronic spins. At high fields ($B<1$~T) the simulated relaxation dynamics, including both inter- 
and intra-spin direct relaxation mechanisms (black curve and dots in Fig.~\ref{tauB2}), shows no significant 
difference from that of two isolated spins relaxing through the modulation of the Zeeman interaction (green 
curve and dots in Fig.~\ref{tauB2}). The relaxation time due to the sole dipolar contribution (red curve and 
dots in Fig.~\ref{tauB2}) becomes predominant only at low fields and it is found to be two orders of magnitude 
slower than that associated to the hyperfine coupling.

\begin{figure}[h!]
\includegraphics[scale=1]{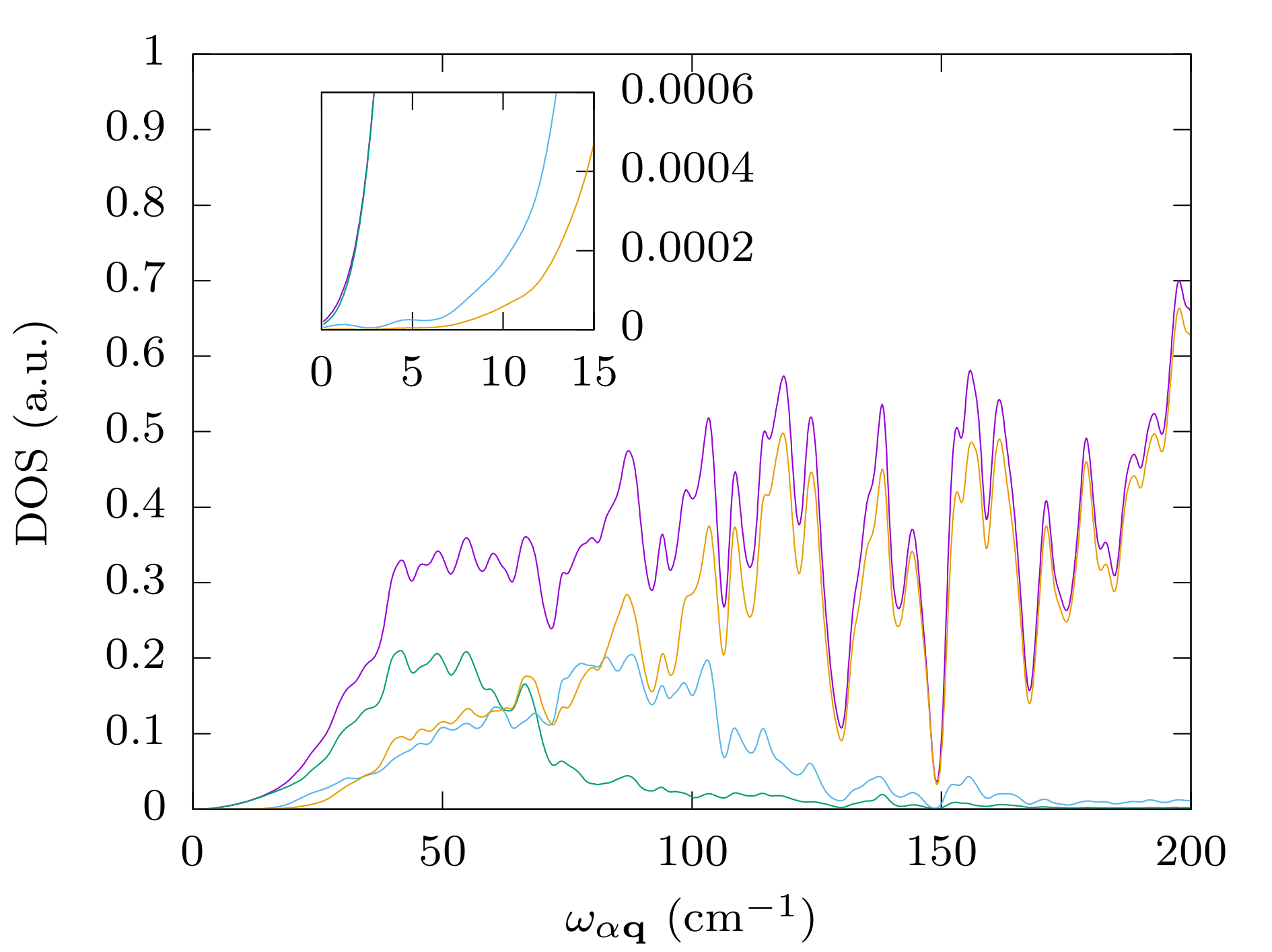}
\caption{\textbf{Phonons density of states.} In purple it is reported the total phonons density of states as a 
function of the frequency. The total phonons density of states has been also decomposed in a pure translational 
contribution (green line), a rotational contribution (blue line) and an intra-molecular contribution (yellow line), all 
relative to a single molecule inside the unit-cell. The inset shows the details of the density of states in the low 
energy part of the spectrum. The Brillouin zone has been integrated with a uniform mesh of $64^{3}$ points. 
The reported density has been smeared with a Gaussian function with breadth 1.0 cm$^{-1}$.}
\label{phdos}
\end{figure}

In order to understand the nature of the interaction between $S=1/2$ spins and the lattice, it is now necessary to 
look at the nature of the phonons involved in the process. It has been shown that, in the absence of spin-spin 
interactions, spin relaxation can only occur through the modulation of the spin Hamiltonian by local rotations 
and intra-molecular distortions~\cite{Lunghi2017a}. However, for a spin 1/2 in reasonable external fields, the 
Zeeman spin splitting (up to a few cm$^{-1}$) is much smaller than the first $\Gamma$-point optical mode, 
here around 50~cm$^{-1}$. Energy conservation [see Eq.~(\ref{delta})] leaves only acoustic phonons as 
candidates for the spin-phonon interaction, suggesting that no energy exchange between lattice and spin is 
possible under these conditions. The solution of this conundrum is provided by the analysis of the phonons nature 
by means of their decomposition into local molecular translation, local molecular rotations and intra-molecular 
distortions~\cite{Lunghi2017a}. The results of this analysis, carried out over the phonons in the entire Brillouin zone, 
are summarized in Fig.~\ref{phdos}, where the total and decomposed phonons density of states are reported.

The total phonons density of states, shows the typical $\sim$$\omega^{2}$ dependence at low frequency, where 
acoustic phonons dominate the vibrational spectra. However, the decomposition of these modes shows that at 
low frequency the nature of the acoustic modes is far from being that of a pure molecular translation. A significant 
rotational and intra-molecular contribution is present at frequencies corresponding to the energy levels' Zeeman 
splitting considered in this work. These contributions provide an efficient relaxation pathway even for a single 
spin isolated from other magnetic centres.

\section{Discussion}

The results of the first-principle calculations presented here are in agreement with Van Vleck's interpretation 
of the spin relaxation in high fields, where a direct relaxation mechanism due to the Land\'e tensor 
modulation is the relevant relaxation pathway. Most importantly we have here demonstrated the mechanism 
underlying the energy transfer between phonons and spins. The presence of an intra-molecular contribution 
inside the low-energy acoustic phonons is of fundamental importance in order to open a relaxation channel, that 
otherwise would be completely inactive due to the translational invariance of the spin-orbit coupling interaction. 

The presence of intra-molecular components in acoustic modes can be understood as a mixing between rigid 
reticular translations and soft molecular modes and it can be used as a rationale to engineer solid-state qubits. 
More rigid molecular modes are expected to diminish sensibly the contamination of low-energy modes, therefore, 
extending the spin lifetime. Such a synthetic strategy has been recently attempted~\cite{Atzori2018,Yamabayashi2018} 
on the basis of similar observations for high-spin molecular magnets~\cite{Lunghi2017a}.  

To conclude, we have presented a general and fully first-principles method to study spin-lattice 
dynamics in magnetic materials. We have predicted the correct spin relaxation time and field dependence for a 
solid-state molecular qubit in high external fields. In particular, we have given, for the first time, a full 
microscopic rationale of the spin relaxation in solid-state molecular qubits by ranking the three fundamental 
interactions at play among electronic and nuclear spins at the first order of perturbation theory. Details concerning 
the spin dynamics at low fields and the nature of the polynomial dependence of the relaxation rate at 
high-$T$~\cite{Tesi2016} remain elusive, suggesting the presence of higher-order processes taking place. The 
method presented here can readily been extended to include higher-order processes such as Raman relaxation 
mechanism and phonon-phonon interactions. This will be the subject of future work.

\section{Methods}
\subsection{Lattice Dynamics}
All the structural optimisation and Hessian calculations have been performed with the CP2K 
software~\cite{VandeVondele2005} at the level of density functional theory (DFT) with the PBE 
functional including Grimme's D3 van der Waals corrections~\cite{Perdew1996,Grimme2010a}. 
A double-zeta polarised (DZVP) MOLOPT basis set and a 600~Ry of plane-wave cutoff have been 
used for all the atomic species. The comparison between simulated and experimental lattice parameters is available in ESI.
All the translational symmetry independent force constants have been 
computed by finite difference approach with a 0.01~\AA\ step. Being $\Phi_{ij}(lm)$ the force constant, 
coupling the $i$-th atomic degrees of freedom in the lattice cell at the position $R_{l}$ and the $j$-th 
atomic degrees of freedom in the lattice cell of position $R_{m}$, the dynamical matrix, 
$\mathbf{D}(\mathbf{q})$, at the \textbf{q}-point, is built as
\begin{equation}
D_{ij}(\mathbf{q})=\sum_{l}\Phi_{ij}^{l0}e^{i\mathbf{q}\cdot\mathbf{R}_{l}}\:.
\end{equation}
The eigenvalues of $\mathbf{D}(\mathbf{q})$ are $\omega^{2}(\mathbf{q})$, while the eigenvectors 
$\mathbf{L}(\mathbf{q})$ defines the normal modes of vibration. The calculated phonons are in good agreement with those previously performed at the sole $\Gamma$-point\cite{Tesi2016}.

\subsection{Spin-Phonon Coupling Coefficients}
The ORCA software~\cite{Neese2012} has been employed for the computation of the $\mathbf{g}$ and 
$\mathbf{A}$ tensors for both equilibrium and distorted geometries. We have used the basis sets def2-TZVP 
for V and O, def2-SVP for C and H and a def2-TZVP/C auxiliary basis set for all the elements. For the 
calculation of the $\mathbf{A}$ tensors the entire basis set have been de-contracted. The calculations of the
$\mathbf{g}$ tensors have been carried out at the CASSCF+NEVPT2 level of theory, with a (1,5) active space 
and spin-orbit contributions included through quasi-degenerate perturbation theory. The calculations of the
$\mathbf{A}$ tensors have been performed at the DFT level with the PBE functional~\cite{Perdew1996}.

The spin phonon coefficients relative to the $\mathbf{g}$ and $\mathbf{A}$ tensors have been calculated 
as numerical derivatives. Ten Cartesian displacements ranging from $\pm 0.01$ \AA.
have been used to estimate $\partial \mathbf{g}/\partial X_{is}$ and $\partial \mathbf{A}/\partial X_{is}$, 
where $X_{is}$ refers to the $s$ Cartesian component of the $i$-th atom in the DFT optimized unit-cell. 
The $\mathbf{g}$ vs $X_{is}$ and $\mathbf{A}$ vs $X_{is}$ profiles have been fitted with a fourth order 
polynomial expression and set to zero if the fitting error on the linear term exceeded 7\%. The spin-phonon 
coupling coefficients relative to the point-dipole-dipole interaction have been obtained by analytical 
differentiation. The Cartesian derivatives $\partial \hat{H}_\mathrm{s}/\partial X_{is}$ have then projected onto the 
normal modes by means of the expression
\begin{equation}   
\Big(\frac{\partial H_\mathrm{s}}{\partial Q_{\alpha\mathbf{q}}}\Big)=\sum_{l}^{N_{cells}}\sum_{is}^{N,3}\sqrt{\frac{\hbar}{N_{q}\omega_{\alpha\mathbf{q}}m_{i}}}e^{i\mathbf{q}\cdot \mathbf{R}_{l}} L^{\alpha\mathbf{q}}_{is} \Big(\frac{\partial H_\mathrm{s}}{\partial X^{l}_{is}}\Big)\:,
\end{equation}
where $X^{l}_{is}$ is the $s$ Cartesian coordinate of the i-th atom of $N$ with mass $m_{i}$, inside the unit-cell 
replica at position $\mathbf{R}_{l}$, and $N_{q}$ is the number of $q$-points used. All the data regarding spin Hamiltonian parameters and their differentiation are reported in ESI.
The spin-phonon coupling distributions have been calculated starting from the spin-phonon coupling coefficients squared norm, defined as
\begin{equation}   
V_{\mathrm{sph}}^{2}(\omega_{\alpha})=\frac{1}{N_{q}}\sum_{\mathbf{q}}\sum_{ij} \Big(\frac{\partial g_{ij}}{\partial Q_{\alpha\mathbf{q}}}\Big)^{2}\:,
\end{equation}
and analogously for the $\mathbf{A}$ and $\mathbf{D}^{\mathrm{Dip}}$ tensors. The Dirac's Delta function appearing in Eq.~(\ref{delta}) has been 
evaluated as a Gaussian function in the limit for infinite $q$-points and vanishing Gaussian breadth. A grid 
of $64^3$ q-points and a Gaussian breadth of 1~cm$^{-1}$ was estimated to accurately reproduce this limit 
for all the temperature and field values investigated. Some results about these convergence tests are reported in ESI. \\

\noindent
\textbf{Supplementary Information}\\
Supplementary information is available: the entire form of the non-secular Redfield equations, comparison between experimental and calculated spin Hamiltonian parameters and crystallographic parameters, the spin Hamiltonain parameters for all molecular distortions and relaxation time convergence tests.\\

\noindent
\textbf{Data Availability}\\
All the relevant data discussed in the present paper are available from the authors upon request. \\

\noindent
\textbf{Acknowledgments}\\
This work has been sponsored by Science Foundation Ireland (grant 14/IA/2624). Computational 
resources were provided by the Trinity Centre for High Performance Computing (TCHPC) and the 
Irish Centre for High-End Computing (ICHEC). We acknowledge Dr. Lorenzo Tesi and Prof. Roberta Sessoli 
for providing the original experimental data and the stimulating scientific discussions. We also acknowledge 
the MOLSPIN COST action CA15128.\\

\noindent
\textbf{Author contributions}\\
All the authors contributed to the discussion of the results and to the manuscript.\\

\noindent
\textbf{Competing financial interests}\\
The authors declare no competing financial interests.


%

\end{document}


\section*{Supplementary Note 1. Non-Secular Redfield Equations.}
The master matrix for the non-secular Redfield equations is
\begin{equation}
R^{\alpha\mathbf{q}}_{ab,cd}=\frac{\pi}{2\hbar^{2}}\left[ V^{\alpha\mathbf{q}}_{ac}V^{\alpha\mathbf{q}}_{db}
\Big(G(\omega_{db},\omega_{\alpha\mathbf{q}})+G(\omega_{cd},\omega_{\alpha\mathbf{q}})\Big)-\sum_{j}\left( V^{\alpha\mathbf{q}}_{aj}V^{\alpha\mathbf{q}}_{jc}\delta_{bd} G(\omega_{jc},\omega_{\alpha\mathbf{q}}) + V^{\alpha\mathbf{q}}_{dj}	V^{\alpha\mathbf{q}}_{jb}\delta_{ca}G(\omega_{jd},\omega_{\alpha\mathbf{q}}) \right) \right]\:,
\label{RED0}
\end{equation}

For a spin Hamiltonian containing N terms, \textit{e.g.} $H_{\mathrm{s}}=\sum_{i}H_{i}$, the terms $V_{ac}^{\alpha\mathbf{q}}V_{db}^{\alpha\mathbf{q}}$ appearing in Eq. \ref{RED0} contains all possible cross-product
\begin{equation}
V_{ac}^{\alpha\mathbf{q}}V_{db}^{\alpha\mathbf{q}}=\sum_{ij}\langle a|\frac{\partial H_{i}}{\partial Q_{\alpha\mathbf{q}}}|c\rangle \langle d|\frac{\partial H_{j}}{\partial Q_{\alpha\mathbf{q}}}|b\rangle \:.
\end{equation}
In our implementation we only considered the same-index terms. This correspond to assuming that the modulation of each term occurs independently from the others.

\section*{Supplementary Note 2. Lattice Parameters and Spin Hamiltonian.}

Supplementary Table \ref{cell} reports the calculated and experimental lattice parameters. Taking into account an expected volume reduction due to the absence of temperature expansion in the simulations, the agreement between experimental and calculated ones is excellent. 

\begin{table}[h!]
\centering
\begin{threeparttable}
\begin{tabular}{l|cccccc}
\\
    Model     &  a (\AA) &  b (\AA) & c (\AA) & $\alpha$ ($^{\circ}$) & $\beta$ ($^{\circ}$) & $\gamma$ ($^{\circ}$) \\
\\
\cline{2-7}
\hline
\\
Exp\tnote{a}  & 7.346 & 8.149 & 11.207 & 72.86 & 72.20 & 66.94 \\
\\
Simulated & 7.060 &  7.935 & 11.091 & 74.467 & 72.622 & 68.216  \\
\\
\hline
\end{tabular}
\begin{tablenotes}
\item[a] Taken from Ref.~\cite{Tesi2016}.
\end{tablenotes}
\end{threeparttable}
\caption{\textbf{Lattice parameters for the VO$\mathbf{(acac)_{2}}$ molecular crystal.} The X-ray experimental lattice parameters are compared with those obtained from the periodic DFT optimization of 3x3x3 super-cell.}
\label{cell}
\end{table}

Supplementary Table \ref{SH} reports the comparison between simulated and experimental values of the spin Hamiltonain paramenters. Both symmetry and magnitude of the eigenvalues of Landè and Hyperfine tensors are well reproduced. The $\mathbf{g}$ tensor anisotropy is slightly overestimated. On the contrary, the eigenvalues of the $\mathbf{A}$ tensor are slightly underestimated. The order of magnitude of these interactions is however well reproduced and we believe that these differences between calculated and experimental values cannot significantly bias our results. 

\begin{table}[h!]
\centering
\begin{threeparttable}
\begin{tabular}{l|cccccc}
\\
    Model     & g$_{x}$ &  g$_{y}$ & g$_{z}$ & $| A_{x} |$ (cm$^{-1}$) & $| A_{y} |$  (cm$^{-1}$) & $| A_{z} |$  (cm$^{-1}$) \\
\\
\cline{2-7}
\hline
\\
Exp\tnote{a}  & 1.9845 & 1.981 & 1.9477 & 0.00580 & 0.00627 & 0.01712 \\
\\
Simulated   & 1.9830 & 1.9814 & 1.9274 & 0.00354 & 0.00396 & 0.01396 \\
\\
\hline
\end{tabular}
\begin{tablenotes}
\item[a] Taken from Ref.~\cite{Tesi2016}.
\end{tablenotes}
\end{threeparttable}
\caption{\textbf{Spin Hamiltonian parameters for VO$\mathbf{(acac)_{2}}$.} The spin Hamiltonian experimental parameters are compared with first-principles ones.}
\label{SH}
\end{table}

\clearpage

\section*{Supplementary Note 4. K-points and Phonon's line-width Convergence.}

In this work we investigated the harmonic phonons limit of the spin dynamics. In this limit, the phonons's green function contains a Dirac's delta distribution $\delta(\omega-\omega_{\alpha \mathbf{q}})$ that enforces energy conservation. The evaluation of the phonons' Green function requires two numerical approximations. One concerns the use of a discrete number of k-points, in principles infinite due to the infinite extension of a crystalline solid, and the approximation of the Dirac's delta distribution in terms of a regular function.

A Gaussian function converges to a Dirac's delta for a Gaussian breadth approaching 0. 
\begin{equation}
\frac{1}{\sigma\sqrt{\pi}}e^{-\frac{(\omega-\omega_{\alpha \mathbf{q}})^{2}}{\sigma^{2}}}\xrightarrow{\text{$\sigma \rightarrow 0$}} \delta(\omega-\omega_{\alpha \mathbf{q}})
\end{equation}

However, this is only true assuming an infinitely dense number of states as function of phonons' frequency. Numerically, this requires the number of k-points to be large enough to guarantee an approximately infinite number of states inside the energy window selected by $\sigma$. Therefore, for each value of $\sigma$, spin relaxation time must be first converged with respect to the number of k-points. Then, sigma is gradually reduced until relaxation time is converged with respect to this quantity as well.

\begin{figure}[h!]
\includegraphics[scale=1]{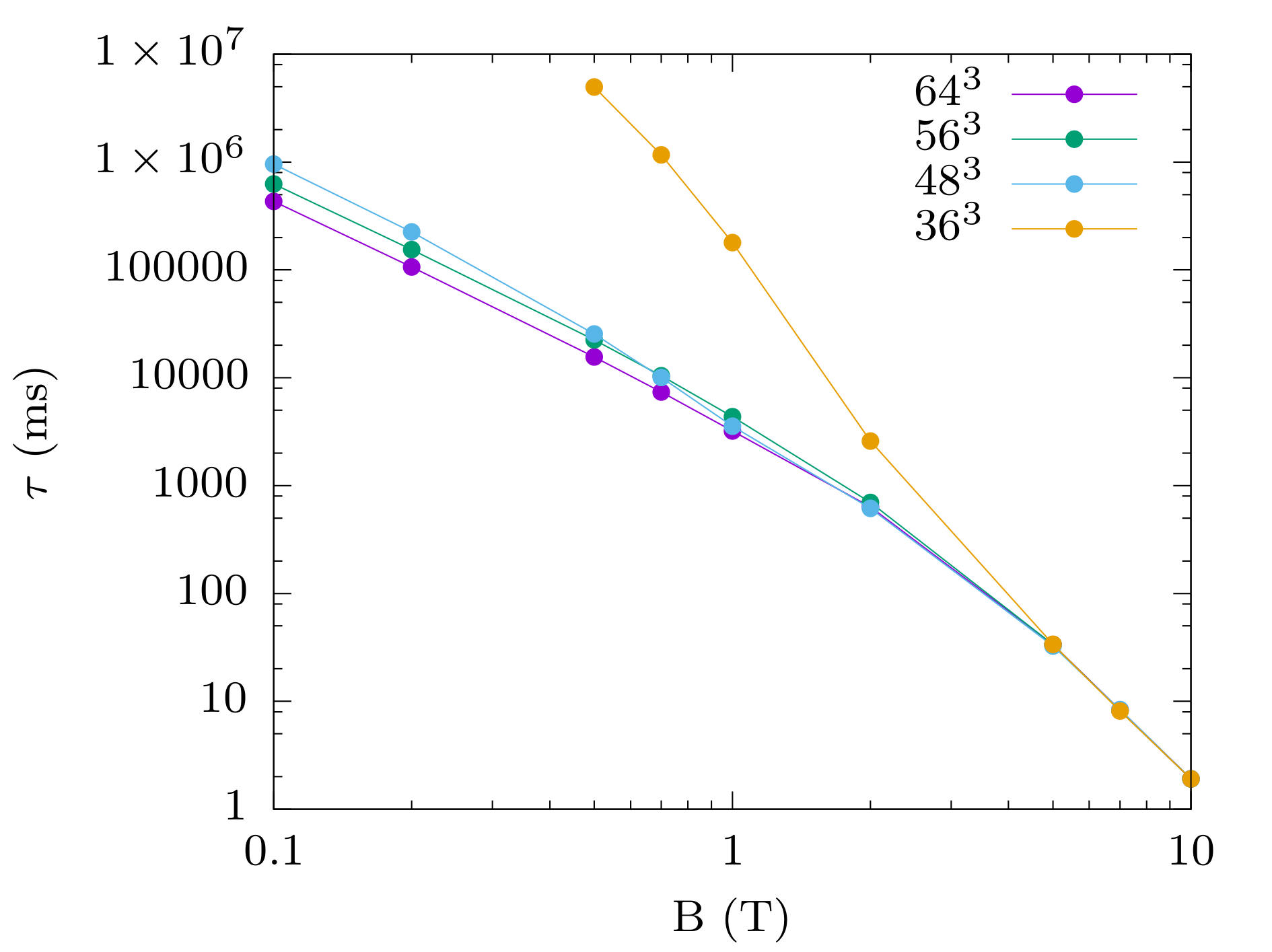}\includegraphics[scale=1]{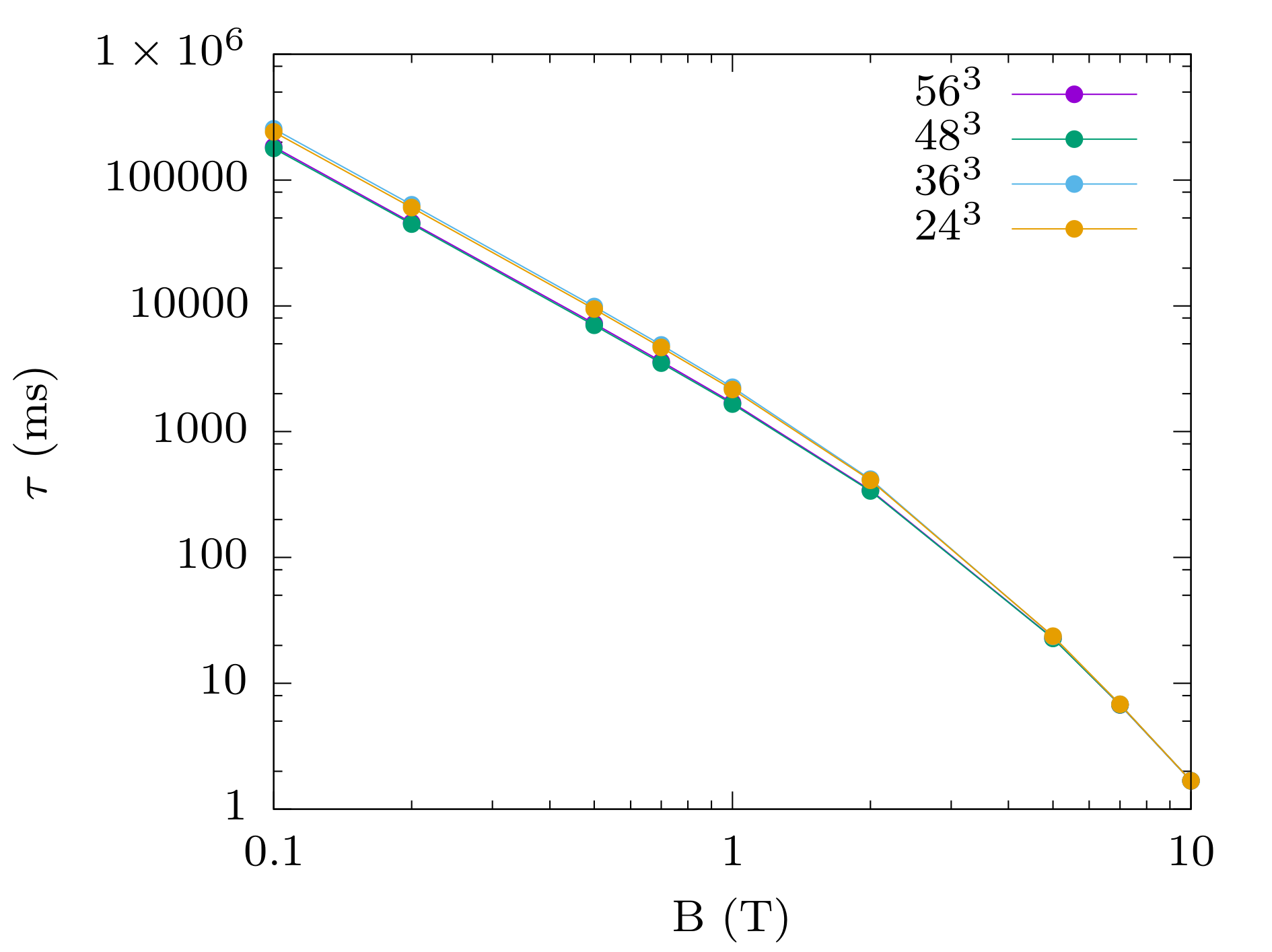}
\caption{\textbf{Spin relaxation time as function of external fields for different values of k-points.} The relaxation time, $\tau$, in ms as function of the Gaussian smearing $\sigma$ for different values of external field magnitude is reported. The simulation corresponds to the dynamics of one electronic spin relaxing due to the phonons modulation of the Zeeman term of the spin Hamiltonian. The left panel correspond to simulations with $\sigma=1$ cm$^{-1}$ and the right Panel to simulations with $\sigma=4$ cm$^{-1}$.}
\label{taukk}
\end{figure} 
The left panel of Supplementary Figure \ref{taukk} shows the convergence of $\tau$ with respect to the number of k-points when $\sigma$ is fixed to 1 cm$^{-1}$. For small Brillouin zone integration grids the spin life-time is largely overestimated. This is because no phonons are available in the near-$\Gamma$ part of the spectra. Increasing the finess of the integration grid the value converges. For larger values of $\sigma$, as reported in the right Panel of Supplementary Figure \ref{taukk}, the convergence is reached for smaller integration grids.

\begin{figure}[h!]
\includegraphics[scale=1]{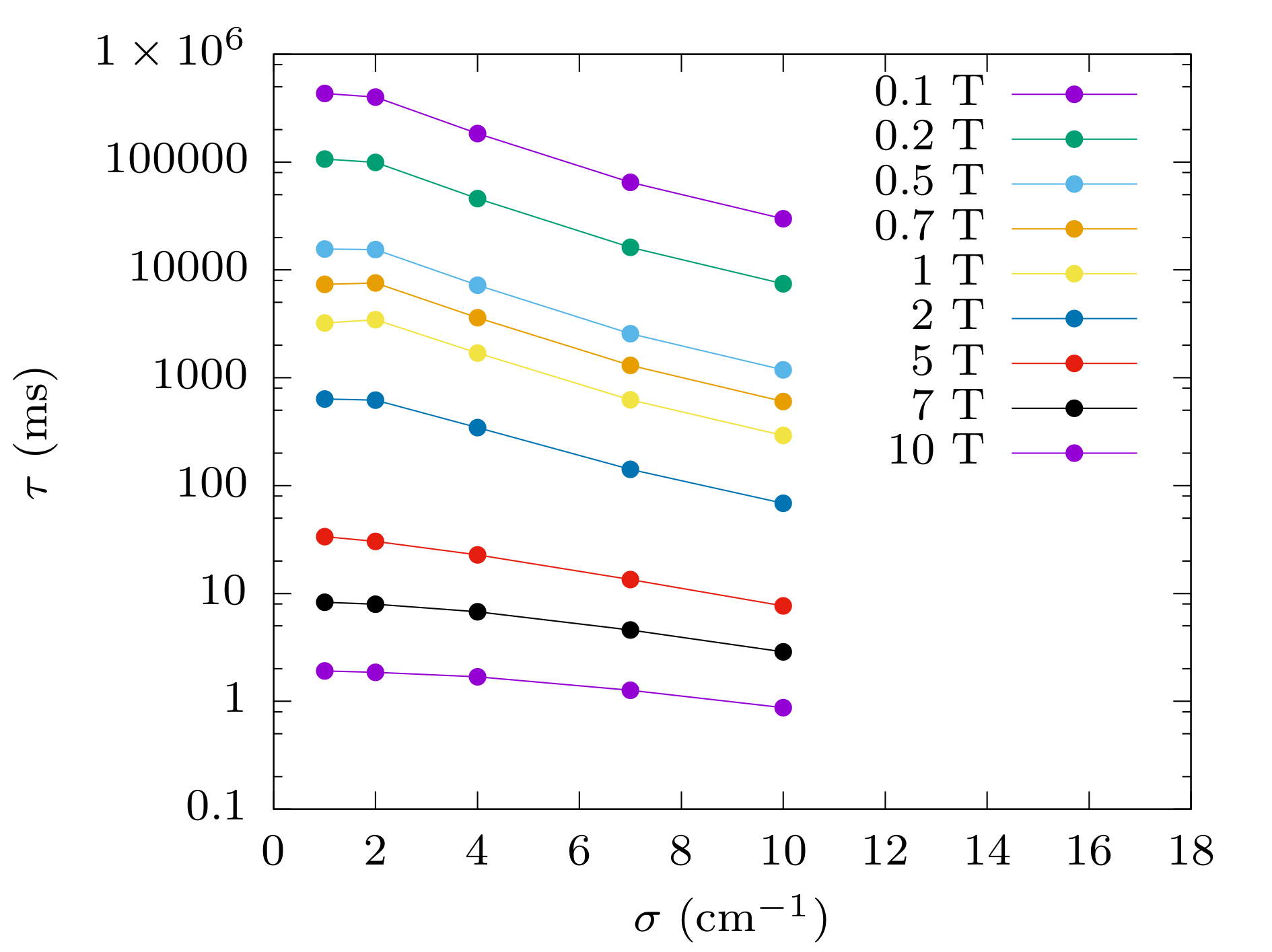}
\caption{\textbf{Spin relaxation time as function of the Gaussian breadth for different external fields.} The relaxation time, $\tau$, in ms as function of the Gaussian smearing $\sigma$ for different values of external field magnitude is reported. The simulation correspond to the dynamics of one electronic spin relaxing due to the phonons modulation of the Zeeman term of the spin Hamiltonian.}
\label{tausigma}
\end{figure} 

Supplementary Figure \ref{tausigma} reports the dependence of the relaxation time $\tau$ with respect to the value $\sigma$. For each value of $\sigma$ the convergence with respect to the number of k-points has been achieved as described above. Relaxation time is nicely converged for $\sigma=1$ cm$^{-1}$. For smaller values of $\sigma$ it was not possible to reach convergence on k-points with a reasonable finess of the Brillouin zone integration grid. This is probably due to the fact that an energy window of less than 1 cm$^{-1}$ is small compared to the numerical noise affecting the phonons calculations.

\clearpage

\section*{Supplementary Note 5. Number of Spins Convergence.}
Due to the exponential increase of the dimension of the master matrix in Eq. \ref{RED0}, the study of relaxation time as function of the number of interacting electronic spins has to be restricted to a maximum of six spins. Within this constraint, six different systems can been studied: two and three unit-cells replicated along x,y and z, respectively. The results are summarized in Supplementary Figure \ref{tausp}. The inclusion of multiple electronic spins beyond the first-neighbour one does not dramatically affect the spin dynamics and the estimations of relaxation time made with just two spins correctly capture the essential physics.
\begin{figure}[h!]
\includegraphics[scale=1]{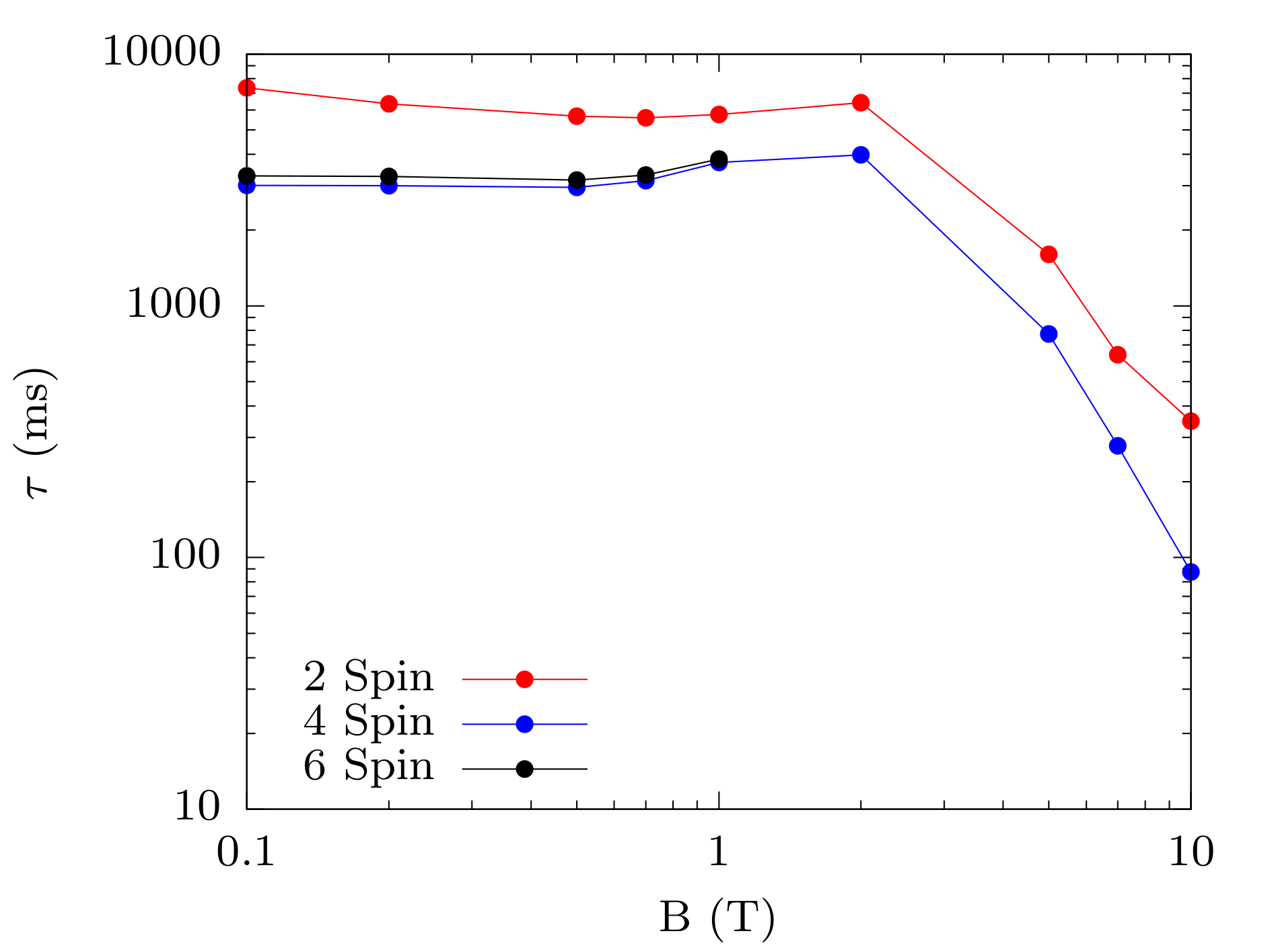}\includegraphics[scale=1]{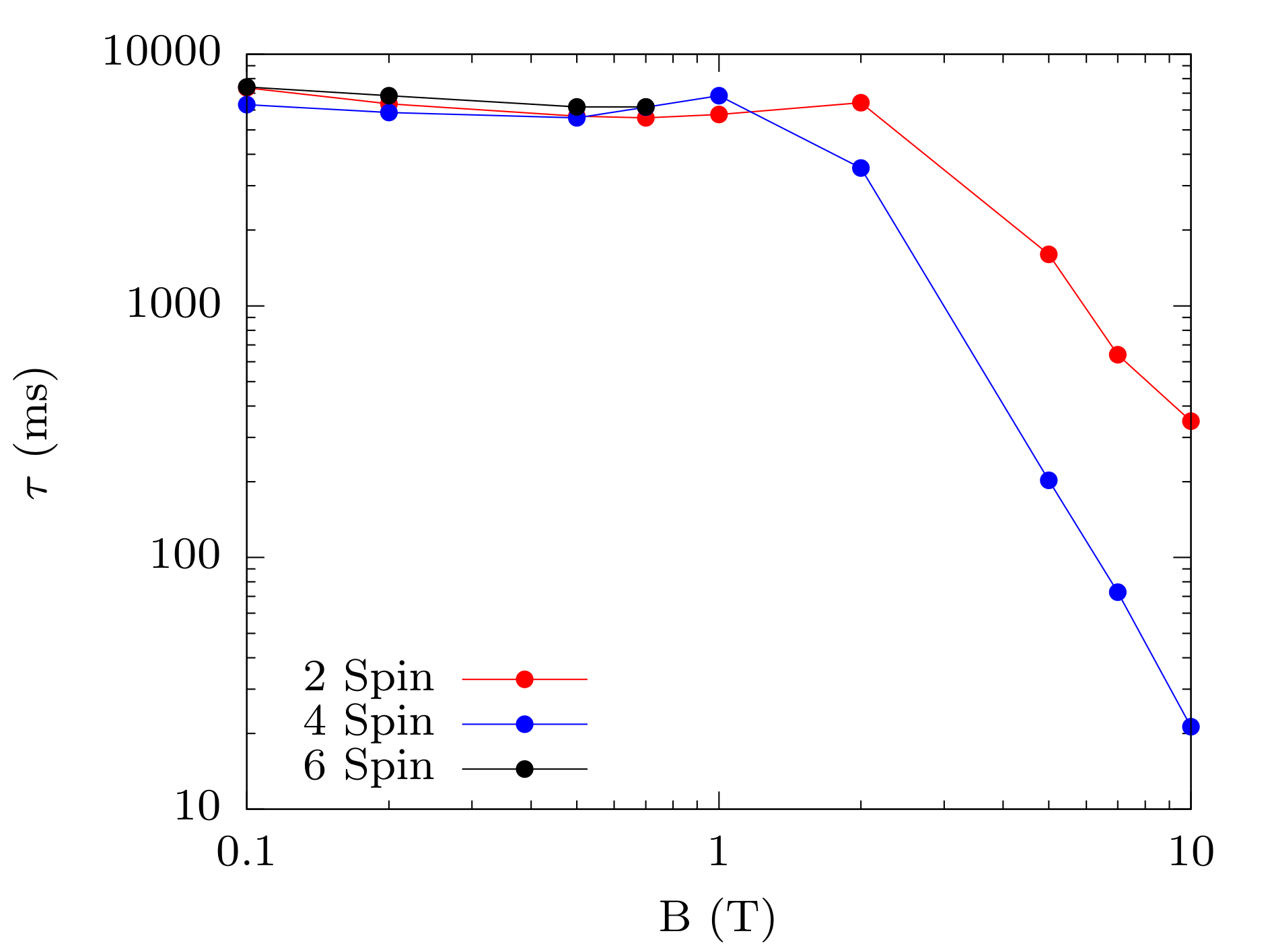}\\
\includegraphics[scale=1]{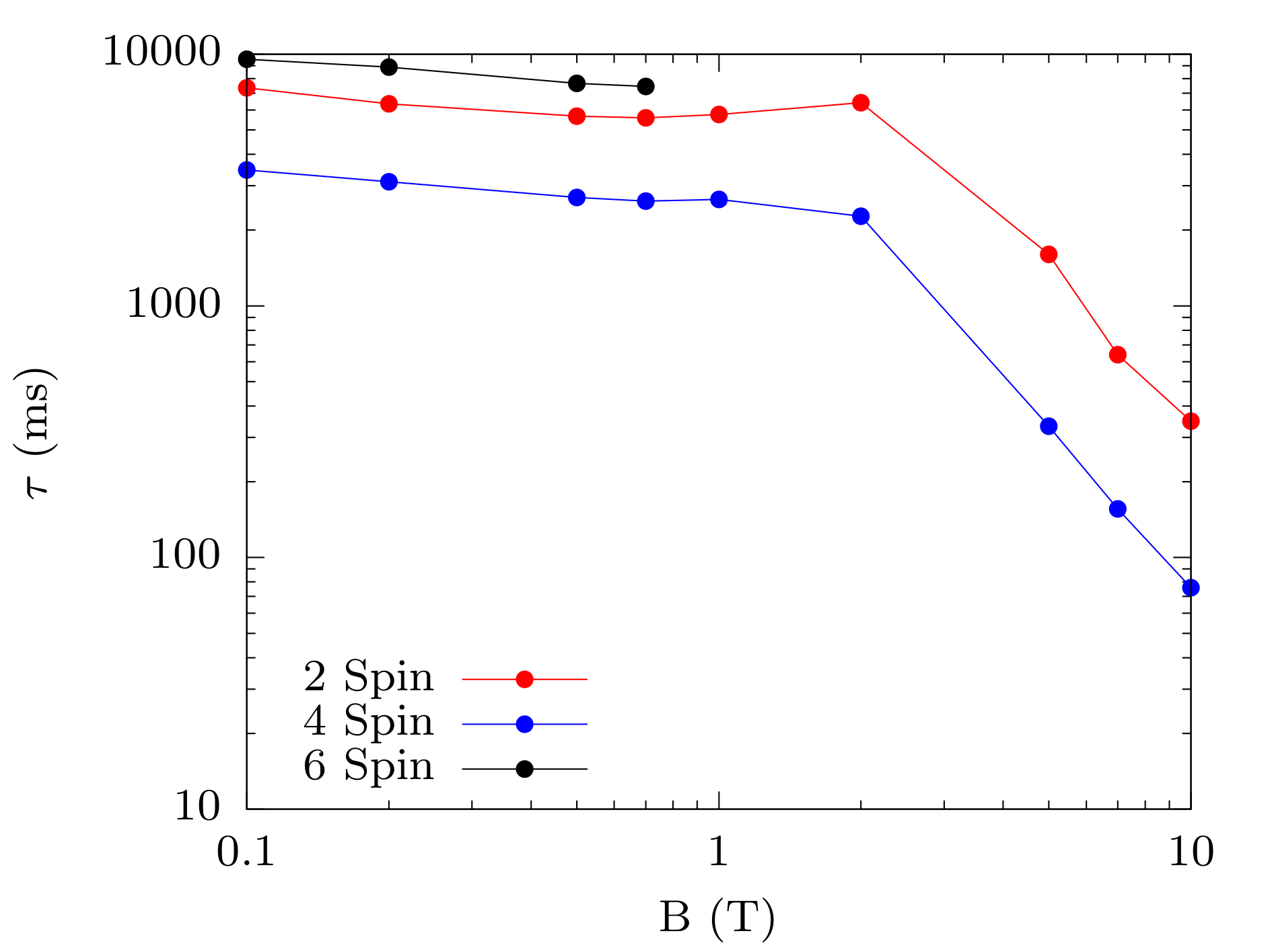}
\caption{\textbf{Spin relaxation time as function of the number of coupled spins.}
The relaxation time, $\tau$, in ms as function of the external field magnitude for different sets of spins is reported: the two spins as in the crystal unit-cell, four spins as obtained by alying two unit-cells and six spins as obtained by alinging three unit-cells. The top-left Panel shows the results for the reticular direction \textbf{a}, top-right Panel shows the results for the reticular direction \textbf{b} and the bottom Panel shows the results for the reticular direction \textbf{c}.}
\label{tausp}
\end{figure} 

\clearpage

\section*{Supplementary Note 6. Effect of Numerical Noise.}

We investigated the effect of numerical noise on the two most important computed quantities: spin-phonon coupling parameters and phonons' frequencies. 
For the former we studied the effect of multiplying by a factor two all the spin-phonon coupling coefficients relatives to the modulation of the hyperfine Hamiltonian, while for the latter we applied a 20\% reduction of all the frequencies. The magnitude of these perturbations has been chosen accordingly to the error these quantities are affected by. In Supplementary Note 2 we have shown that hyperfine parameters are underestimated by a factor of almost two. Comparing our phonons simulations with THz spectroscopy\cite{Atzori2017} we observe an error on the first frequency at the $\Gamma$-point, here calculated at $\sim 50$ cm$^{-1}$, of no more than 20\%.

\begin{figure}[h!]
\includegraphics[scale=1]{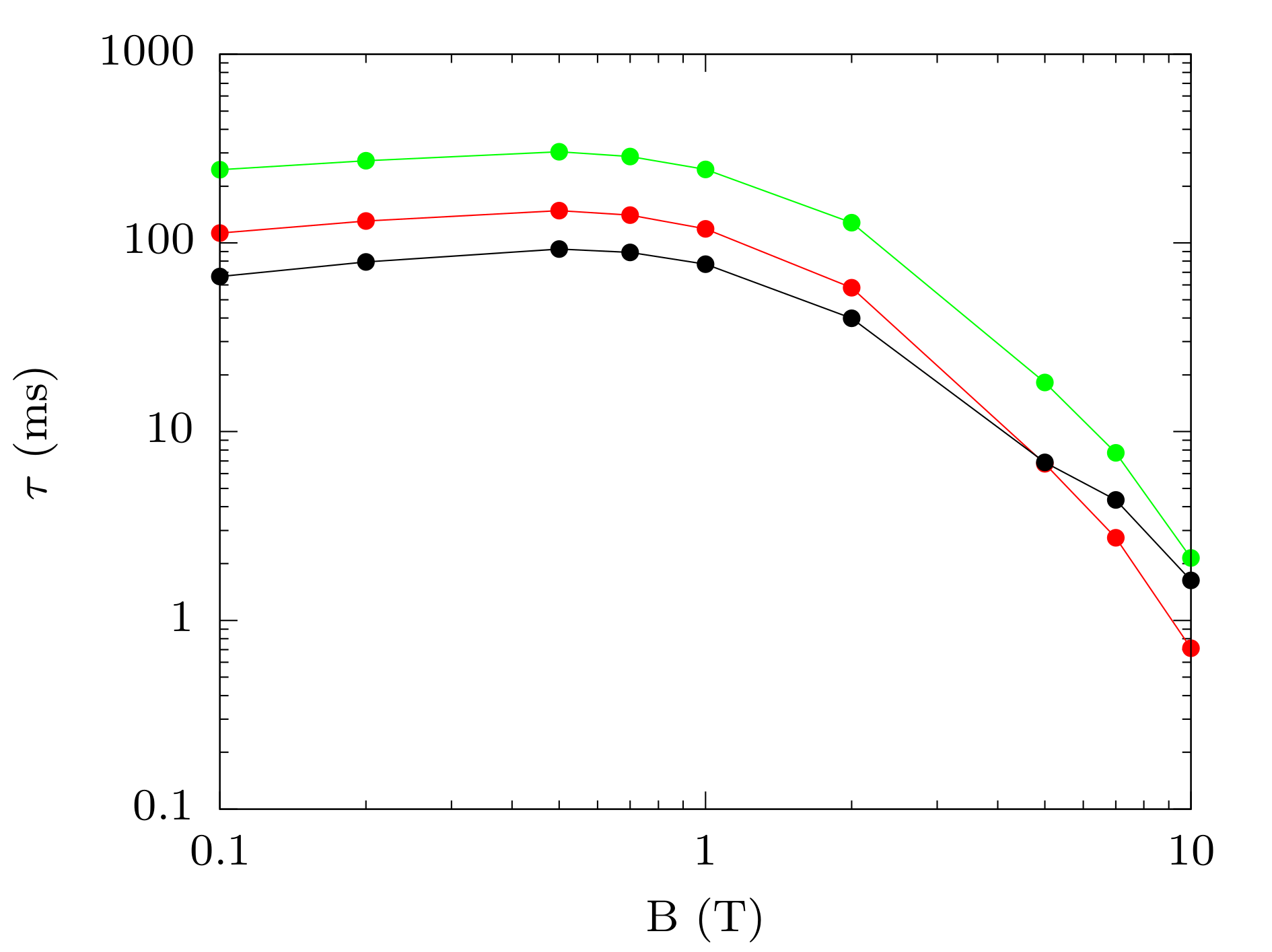}
\caption{\textbf{Effect of numerical noise on spin relaxation time.} The relaxation time, $\tau$, in ms as function of the external field magnitude in Tesla for the unbiased case (Green line and dots), for a double hyperfine spin-phonon coupling (Black line and dots) and for a 20\% rescaled frequencies (Red line and dots).}
\label{noise}
\end{figure} 

Supplementary Figure \ref{noise} reports the comparison between the unbiased dynamics (Green line and dots) and the dynamics in presence of a doubled hyperfine interaction (Black line and dots) and the dynamics obtained after frequencies rescaling (Red line and dots). The effect of both the perturbations investigated show that numerical errors in the determination of spin-phonon coupling coefficients and phonons' frequencies do not significantly affect our results and the orders of magnitude of our estimations are well converged.

\clearpage


%